\documentclass[%
 aip,
 amsmath,amssymb,
 reprint,%
]{revtex4-1}

\usepackage{graphicx}
\usepackage{dcolumn}
\usepackage{bm}

\usepackage[utf8]{inputenc}
\usepackage[T1]{fontenc}
\usepackage{mathptmx}
\usepackage{etoolbox}
\usepackage{xcolor}
\usepackage{orcidlink}
\usepackage{braket}
\makeatletter
\def\@email#1#2{%
 \endgroup
 \patchcmd{\titleblock@produce}
  {\frontmatter@RRAPformat}
  {\frontmatter@RRAPformat{\produce@RRAP{*#1\href{mailto:#2}{#2}}}\frontmatter@RRAPformat}
  {}{}
}%
\makeatother
\begin{document}

\preprint{AIP/123-QED}

\title{Super-Heisenberg Non-Equilibrium Quantum Sensing with Waveguide-Coupled Emitters}
\author{Mohammad B. Arjmandi~\!\!\orcidlink{0000-0001-7046-3818}}
 \email{mohammad.arjmandi@upol.cz}
\affiliation{ 
Department of Optics, Palack\'y University, 17. listopadu 12, 779 00 Olomouc, Czech Republic 
}
\affiliation{Wilczek Quantum Center, Shanghai Institute for Advanced Studies, University of Science and Technology of China, Shanghai 201315, China}

\date{\today}

\begin{abstract}
We explore an array of quantum emitters as non-equilibrium probes, coupled to a one-dimensional photonic waveguide, aiming to estimate its properties such as wave number which encodes the waveguide’s frequency and dispersive characteristics. By considering transient dynamics following initial excitation, we show that the quantum Fisher information (QFI) can be significantly enhanced through careful emitter positioning. For two-emitter probes, optimal spacing stabilizes populations and coherences in the single-excitation subspace, suppressing superradiant decay and extending both the magnitude and longevity of QFI. Randomized emitter configurations also reveal that vanishing waveguide-mediated cross decay maximizes both achievable sensitivity and the temporal duration over which information about the parameter remains accessible. Extending to multipartite probes, we demonstrate that the maximum QFI and its temporal integral scale with system size, exceeding the Heisenberg limit for all positioning strategies. Our results highlight the potential of waveguide-coupled emitter arrays as versatile quantum sensors, where collective radiative dynamics can be harnessed to achieve tunable, long-lived, and enhanced precision. 
\end{abstract}

\maketitle


\section{\label{sec:Introduction}Introduction}
Quantum metrology leverages nonclassical features to estimate physical parameters with precision beyond what is attainable using any strategy that relies solely on classical resources~\cite{giovannetti2011advances,giovannetti2006quantum,toth2014quantum}. Over the past decades, quantum-enhanced metrology has developed into a central field of quantum technologies, with applications ranging from estimation of time and frequency~\cite{rosenband2008frequency,bollinger1996optimal,huelga1997improvement,nichol2022elementary,schulte2020prospects} and phase~\cite{vidrighin2014joint,braunstein1992quantum,giovannetti2004quantum} to magnetometry~\cite{schmitt2017submillihertz,jones2009magnetic,zheng2023entanglement,loretz2014nanoscale,maze2008nanoscale}, gravitational-wave detection~\cite{schnabel2010quantum,zhang2023gravitational,kolkowitz2016gravitational,graham2013new,khalili2018overcoming,tse2019quantum}, and imaging~\cite{morris2015imaging,barreto2022quantum,casacio2021quantum}.

In classical parameter estimation, the uncertainty in measuring a parameter decreases only as $1/\sqrt{N}$, where $N$ is the total exploited resources, such as the number probes or independent measurement events. This scaling, commonly known as the standard quantum limit (SQL), originates from the fact that each probe contributes independent fluctuations, which accumulate in form of the so-called shot noise~\cite{demkowicz2015quantum,paris2009quantum,degen2017quantum}. Surpassing the SQL requires exploiting genuine quantum resources. For instance, squeezed states redistribute quantum fluctuations by reducing uncertainty in one conjugate quadrature, to enhance phase sensitivity~\cite{ligo2011gravitational,aasi2013enhanced,oelker2016audio}. Entanglement also enables sensitivities beyond classical bounds, a prominent example being the family NOON states which are maximally path-entangled structures~\cite{jones2009magnetic,dowling2008quantum}. Likewise, Greenberger-Horne-Zeilinger (GHZ) states—exhibiting genuine multipartite entanglement across several constituents—enable Heisenberg limit of sensitivity, in which the uncertainty in estimating a parameter scales as $1/N$, surpassing the SQL~\cite{giovannetti2004quantum,leibfried2004toward}. Despite these advantages, preparing, stabilizing and readout of such highly entangled states remains experimentally demanding due to their extreme fragility to noise and decoherence~\cite{dur2004stability,aolita2008scaling,lu2014experimental,nolan2017optimal}.

Precision enhancement can also arise in many-body quantum systems, where interactions and collective dynamics provide powerful metrological resources~\cite{montenegro2025quantum,baak2024self,manshouri2025quantum}. These systems often generate useful entanglement intrinsically and can exhibit robustness against ubiquitous noises. Quantum criticality, in particular, enables strong amplification of parameter sensitivity through diverging susceptibility around phase transitions points~\cite{salvia2023critical,he2023stark,sarkar2022free,montenegro2023quantum}.

Although dissipation is usually expected to degrade the quantum resources that enable high-precision sensing, reservoir engineering can in fact enhance parameter estimation by tailoring probe-reservoir interactions~\cite{sha2022continuous}, or by dissipatively generating squeezed states~\cite{xia2023time,groszkowski2022reservoir}, thereby eliminating the need for externally prepared squeezed inputs.

On the other hand, another route to engineer dissipative interactions is provided by coupling quantum emitters to the guided modes of a one-dimensional photonic waveguide. In such systems, the interplay of coherent and dissipative dynamics gives rise to collective phenomena, including superradiant decay channels, long-lived subradiant states, and tunable dipole–dipole interactions~\cite{albrecht2019subradiant,cardenas2023many}. These effects have been extensively explored in recent theoretical and experimental studies. For instance, asymmetric driving of a homogeneous atomic array can lead to strongly confined single-site excitation localization~\cite{chung2025strongly}, while selectively radiant states in ordered arrays allow enhanced photon storage and controlled emission into desired channels~\cite{asenjo2017exponential}. Additionally, waveguide-coupled emitters can exhibit many-body localization in the presence of positional disorder~\cite{fayard2021many}. Importantly, the properties of the collective modes—such as decay rates and interaction strengths—are highly sensitive to the precise positions of the emitters, which naturally motivates the question of whether such systems can be exploited as quantum probes to extract information about the waveguide itself or other parameters encoded in the emitter–waveguide interface.

Building on these insights, in this work we investigate whether arrays of quantum emitters coupled to a one-dimensional waveguide can  serve as sensitive probes of the waveguide’s properties. In particular, we employ the emitters to estimate the wave number of the guided mode—a parameter directly linked to the waveguide’s dispersion, refractive index, and frequency response. Our approach focuses on a non-equilibrium sensing strategy, in which the probe evolves solely via spontaneous emission and is interrogated during its transient decay dynamics, long before it relaxes to the trivial steady state, which carries no information about the parameter of interest. Each emitter undergoes both collective, waveguide-mediated decay into the guided mode and local, independent decay into non-guided free-space modes. To avoid introducing any initial quantum resources whose preparation may itself be demanding, we initialize the emitters in their pure excited state, ensuring that any metrological advantage arises solely from the dissipative dynamics mediated by the waveguide. In contrast to conventional steady-state sensing protocols, where the parameter is extracted from a stationary driven or thermalized state, the present waveguide-QED setup retains useful metrological information primarily during the transient relaxation dynamics. As the probe continuously loses excitation through guided and nonguided decay channels, the long-time steady state becomes essentially insensitive to the waveguide parameter, making it advantageous to interrogate the system before the encoded information is completely erased. We first analyze the single-emitter probe and show that the quantum Fisher information (QFI) and its lifetime can be markedly enhanced, through proper positioning of the emitter relative to the guided mode. This enhancement originates from the ability to suppress the effective waveguide-induced decay rate, thereby preserving information about the wave number for longer times. 

We then extend the analysis to two-emitter probe, where appropriate positioning again leads to substantial QFI enhancement. This effect is directly linked to the dynamics of the single-excitation subspace, for which carefully chosen inter-emitter spacing stabilizes the populations and coherences associated with collective decay channels. In particular, suppressing the superradiant decay rate allows the QFI to grow initially, reach a larger maximum, and persist over extended times. Studying ensembles with randomized positions, we find that both the peak of QFI and the integrated QFI over time which measures its durability, are maximized when the waveguide-mediated cross decay vanishes, effectively suppressing the superradiant channel and stabilizing the estimation precision. 

Finally, we generalize our findings to multipartite probes. We demonstrate that both the maximal QFI and its temporal integral scale favorably with the number of emitters, and, strikingly, that for all spatial configurations the scaling of the maximum QFI surpasses the Heisenberg limit. This shows that collective waveguide-mediated dissipation can be harnessed not only to preserve metrological information but also to generate super-Heisenberg precision scaling without any initial entanglement or specially prepared quantum resources.

We note that super-Heisenberg scaling has previously been discussed in several quantum-metrological settings, including nonlinear many-body interactions and interaction-induced parameter encoding~\cite{boixo2007generalized,napolitano2011interaction}, localization- and criticality-enhanced sensing~\cite{he2023stark}, as well as driven nonequilibrium phases and Floquet-engineered quantum probes~\cite{lyu2020eternal,mishra2021driving}. In contrast to these approaches, the enhancement mechanism identified here originates from geometry-controlled waveguide-mediated dissipation during transient non-equilibrium evolution. In particular, the observed scaling emerges dynamically from collective radiative interference and spatially engineered emitter-waveguide interactions, without relying on the mechanisms employed in those previous settings.

\section{Quantum Sensing Framework}
\label{Quantum Sensing Framework}
Quantum sensing aims to exploit unique quantum resources like entanglement to achieve precision in parameter estimation beyond classical limits. The general strategy consists of preparing a well-controlled quantum system, namely the probe, that interacts with another system or environment of interest in such a way that the parameter to be estimated, denoted by $X$, becomes encoded in the probe state through a suitable dynamical process~\cite{paris2009quantum,barbieri2022optical,huang2024entanglement}. This encoding process can be represented in full generality by a completely positive trace-preserving (CPTP) map,
\begin{equation}
\rho(X) = \mathcal{E}_X[\rho_0],
\end{equation}
where $\rho_0$ is the initial state of the probe and $\mathcal{E}_X$ describes the evolution that imprints the information about the parameter $X$ onto the probe. This process does not need to be unitary and may include dissipation or dephasing effects, depending on the nature of the interaction between the probe and its environment.

Once the parameter is encoded, a measurement is performed on the probe, described by a positive operator-valued measure (POVM) $\{ M_m \}$, yielding outcome probabilities 
\begin{equation}
p(m|X) = \mathrm{Tr}[M_m \rho(X)].
\end{equation}
The measurement data are then processed statistically to infer the value of $X$, typically using an unbiased estimator $\hat{X}$ whose variance quantifies the precision of the sensing protocol. For any classical measurement strategy, the Cramér–Rao bound~\cite{cramer1999mathematical,rao1945information} sets a lower limit to the achievable uncertainty:
\begin{equation}
\mathrm{Var}(\hat{X}) \ge \frac{1}{F_C(X)},
\end{equation}
where $F_C(X)$ is the classical Fisher information associated with the probability distribution $\{p(m|X)\}$ given by
\begin{equation}
F_C(x) = \sum_m \frac{\big[\partial_X p(m|X)\big]^2}{p(m|X)}.
\end{equation}
Maximizing $F_C(X)$ over all possible measurements leads to the quantum version of Fisher information~\cite{braunstein1994statistical}, $F_Q(X)$, which quantifies the ultimate precision attainable in estimating $X$ given the quantum state $\rho(X)$:
\begin{equation}
F_Q(X) = \max_{\{M_m\}} F_C(X) = \mathrm{Tr}\!\left[ \rho(X) L_X^2 \right],
\end{equation}
where $L_X$ is the symmetric logarithmic derivative (SLD), implicitly defined by 
\begin{equation}
\partial_X \rho(X) = \frac{1}{2}\left( L_X \rho(X) + \rho(X) L_X \right).
\end{equation}
The corresponding quantum Cramér–Rao bound reads~\cite{braunstein1996generalized}
\begin{equation}
    \mathrm{Var}(\hat{X}) \ge \frac{1}{F_Q(X)}.
\end{equation}

For a general mixed state, the QFI depends on both the eigenvalues and eigenvectors of $\rho(x)$. However, in the special case of a pure state $\rho(X) = \ket{\psi(X)}\bra{\psi(X)}$, it simplifies to~\cite{fujiwara1995quantum}
\begin{equation}
F_Q(X) = 4\left[ \braket{\partial_X \psi(X) | \partial_X \psi(X)} - |\braket{\psi(X) | \partial_X \psi(X)}|^2 \right].
\end{equation}

This formalism provides a universal framework for quantum parameter estimation, irrespective of whether the underlying dynamics are unitary or dissipative. In the present work, we employ this framework to evaluate the sensitivity of an array of emitters coupled to a semi-infinite waveguide, where the parameter of interest is the effective wave number of the guided mode.

\section{Model}
\label{Model}
In this section, we introduce the physical model and the underlying dynamics of our system. We consider an array of identical two-level systems (TLSs) acting as quantum emitters, each characterized by a ground state $\ket{\rm g}$, an excited state $\ket{\rm e}$, and a resonant transition frequency $\omega_0$. The emitters are positioned at fixed locations $\{z_j\}$ along a one-dimensional semi-infinite photonic channel, namely, a waveguide, which supports a continuum of electromagnetic modes~\cite{goban2014atom,goban2015superradiance,hood2016atom,van2013photon,hoi2011demonstration,mirhosseini2019cavity}. The interaction between the emitters and the guided modes gives rise to rich collective radiative behavior and serves as the basis for using the system as a quantum probe.

The total Hamiltonian of the system can be written as
\begin{equation}
H = H_{\mathrm{em}} + H_{\mathrm{w}} + H_{\mathrm{int}},\label{total_Hamiltonian}
\end{equation}
in which the first term, $H_{\mathrm{em}}$, describes the internal energy of the emitters, while $H_{\mathrm{w}}$ accounts for the energy of the photonic continuum and the last term, $H_{\mathrm{int}}$, governs the interaction between the emitters and the waveguide modes. Each term is given by ($\hbar=1$)
\begin{align}
H_{\mathrm{em}} =  \omega_0 \sum_{j} \sigma^{\dagger}_{j} \sigma_{j},\label{TLS_Ham}
\end{align}
\begin{align}
H_{\mathrm{w}} =  \sum_{\nu = \pm} \int \omega_{k} \, b^{\dagger}_{\nu,k} b_{\nu,k} \, dk,\label{modes_Ham}
\end{align}
\begin{align}
H_{\mathrm{int}} =  g \sum_{j,\nu = \pm} \int \!\! \left( b_{\nu,k} \, \sigma^{\dagger}_{j} e^{i \nu k z_j} + \text{H.c.} \right) dk,\label{interaction_Ham}
\end{align}
Here, $\sigma^{\dagger}_{j}$ ($\sigma_{j}$) are the raising (lowering) operators of the $j$-th emitter, and $b_{\nu,k}^{\dagger}$ ($b_{\nu,k}$) denote the creation (annihilation) operators of the photonic mode with wavevector $k$ and frequency $\omega_k$, propagating in the direction $\nu = \pm$ along the waveguide. The coupling constant $g$ represents the uniform light–matter interaction strength between each emitter and the guided field modes. Owing to the one-dimensional geometry and the semi-infinite boundary conditions, interference between left- and right-propagating modes can significantly modify the collective emission dynamics of the emitter array.
To describe the dynamics of the emitters, we move to the interaction picture with respect to the free Hamiltonian of the field and the emitters. We assume that the coupling between the emitters and the waveguide modes is sufficiently weak such that the photonic reservoir remains nearly intact by its interaction with the emitters. Moreover, the correlation functions of the reservoir decay much faster than the characteristic timescale of the emitters’ dynamics. These assumptions lead to the Born–Markov approximation~\cite{breuer2002theory,rivas2012open}, under which the memory effects can be neglected and the reservoir can be treated as a Markovian bath (zero temperature) with no temporal correlations. In addition to dissipation into the guided modes of the waveguide, each emitter can also decay into non-guided modes of the surrounding free space. These processes introduce an additional local, independent decay channel for each emitter. By tracing out the photonic degrees of freedom, we obtain the following master equation for the reduced density matrix $\rho$ of the emitters:
\begin{equation}
\dot{\rho} = -i [H_{\mathrm{1D}}, \rho] 
+ \mathcal{D}_{\rm G}(\rho)+\mathcal{D}_{\rm NG}(\rho),\label{eq:master}
\end{equation}
where the effective coherent interaction between emitters is governed by~\cite{chang2012cavity,lalumiere2013input,albrecht2019subradiant}
\begin{equation}
H_{\mathrm{1D}} = -\sum_{i,j} J_{ij} \, \sigma^{\dagger}_{i} \sigma_{j}.\label{H1D}
\end{equation}
The collective and individual spontaneous emissions, respectively into guided and non-guided modes are given by
\begin{equation}
\mathcal{D}_{\rm G}(\rho)=\sum_{i,j} \Gamma^{\rm s}_{ij} 
\left( \sigma_{i} \rho \sigma^{\dagger}_{j}
- \tfrac{1}{2} \{ \sigma^{\dagger}_{j} \sigma_{i}, \rho \} \right),\label{D_Guided}
\end{equation}
\begin{equation}
\mathcal{D}_{\rm NG}(\rho)=\sum_{i} \gamma_i 
\left( \sigma_{i} \rho \sigma^{\dagger}_{i}
- \tfrac{1}{2} \{ \sigma^{\dagger}_{i} \sigma_{i}, \rho \} \right).\label{D_nonGuided}
\end{equation}
\begin{figure}[t!]
\begin{center}
\includegraphics[width=1\columnwidth]{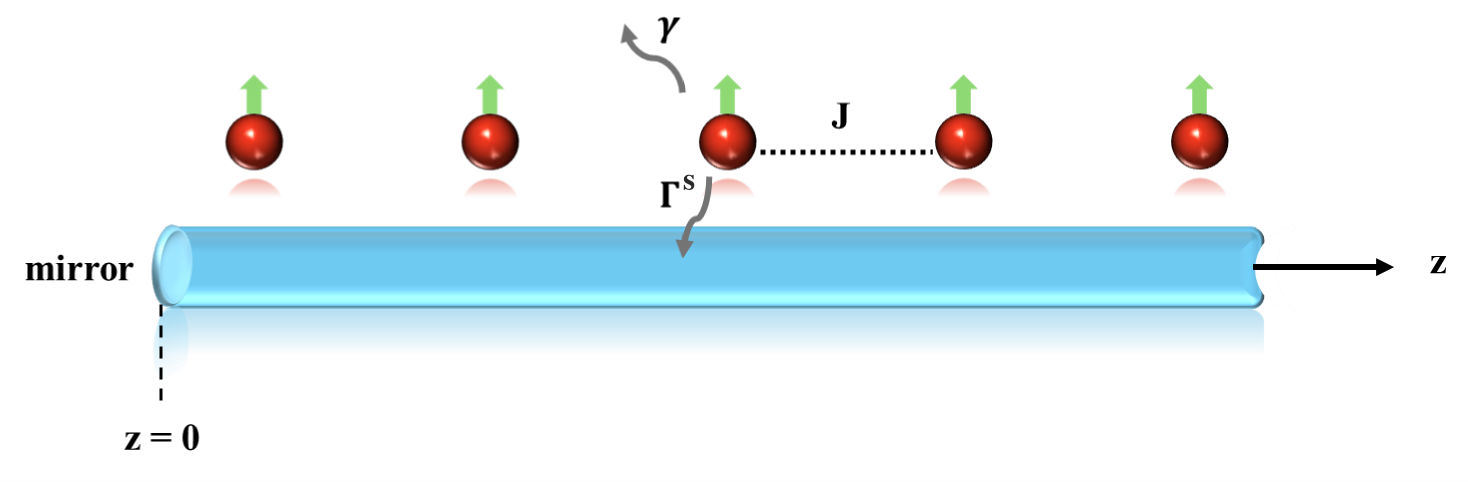}
\end{center}
\caption{Schematic figure of the considered model, including an array of emitters as a quantum probe coupled to a 1D waveguide. The emitters can be evenly spaced or randomly positioned. A highly reflective mirror is placed at one end of the waveguide ($z=0$). The emitters can experience both collective waveguide mediated decay $\Gamma^{\rm s}$ and independent local decay $\gamma$ into free space.}
\label{sketch}
\end{figure}
 To model the semi-infinite waveguide, we place a perfectly reflecting mirror at position $z=0$, as depicted in Fig.~\ref{sketch}, which imposes a boundary condition on the electromagnetic field at one end of the waveguide. As a result, photons propagating toward the mirror are reflected back into the waveguide, effectively creating interference between the incident and reflected fields. This reflection modifies the emitter–emitter interactions and gives rise to additional terms in the coherent and dissipative couplings that depend on both $|z_i - z_j|$ and $|z_i + z_j|$. Thus, the waveguide-mediated coherent and dissipative couplings between emitters are, respectively, given by~\cite{albrecht2019subradiant,cardenas2023many}
\begin{align}
J_{ij} = \frac{\gamma_{\mathrm{1D}}}{2} \left[ \sin\!\left( k_{\mathrm{1D}} |z_i - z_j| \right) - \sin\!\left( k_{\mathrm{1D}} |z_i + z_j| \right) \right],\label{coupling}
\end{align}
\begin{align}
\Gamma^{\rm s}_{ij} = \gamma_{\mathrm{1D}} 
\left[ \cos\!\left( k_{\mathrm{1D}} |z_i - z_j| \right) - \cos\!\left( k_{\mathrm{1D}} |z_i + z_j| \right) \right],\label{decay}
\end{align}
Here, $\gamma_{\mathrm{1D}}$ denotes the spontaneous emission rate of a single emitter into the waveguide modes, and $k_{\mathrm{1D}}$ is the resonant wave number of the guided photons. We also fix the local (unguided) dissipation rate of each emitter to $\gamma$.

In the following sections, we employ the emitters as a quantum probe of the photonic environment, namely, the effective wave number $k_{\mathrm{1D}}$, which in turn provides access to the frequency or dispersion properties of the guided mode. We begin our analysis with the simplest case of a single emitter coupled to the semi-infinite waveguide and then extend the discussion to multiple emitters, where collective interference effects become significant.
\begin{figure*}[t]
\begin{center}
\includegraphics[width=1\textwidth]{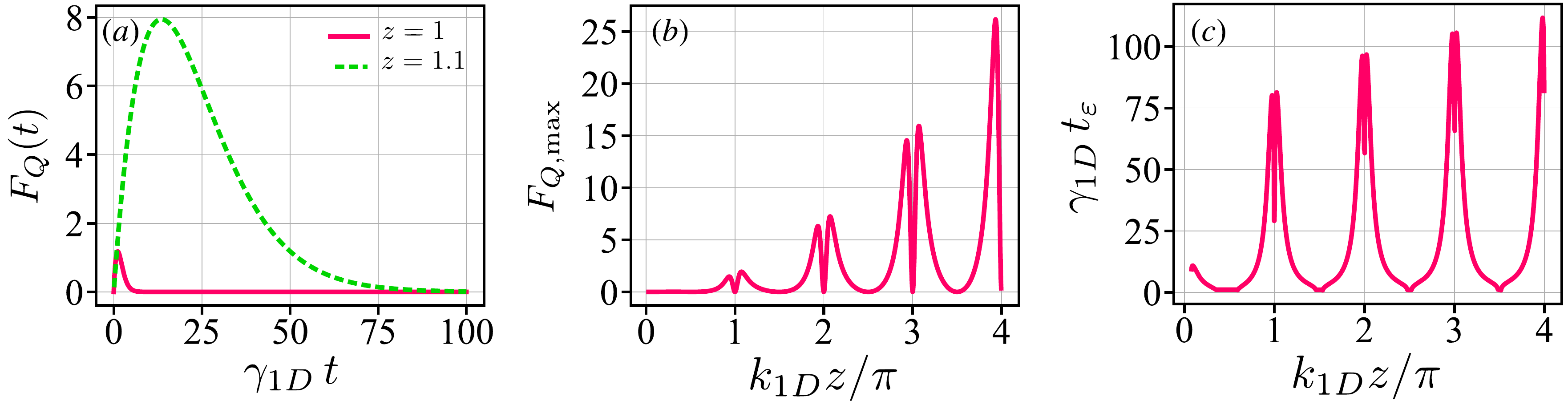}
\end{center}
\caption{(a) Time-dependence of QFI for single emitter probe, given by Eq.~\eqref{qfi_single_qubit} with $k_{1D}=2.7 \pi$, (b) maximum QFI and (c) temporal window over which the QFI remains non zero, against the emitter position, respectively given by~\eqref{single_qubit_qfi_max} and~\eqref{time_epsilon}. The local decay rate is fixed as $\gamma=0.1 \gamma_{1D}$.}
\label{single_qubit}
\end{figure*}
\section{Non-equilibrium Quantum Sensing}
\label{Non-equilibrium Quantum Sensing}
Quantum sensing can operate in two distinct regimes: \emph{equilibrium} and \emph{non-equilibrium}. 
In the equilibrium regime, the probe interacts with its environment until the joint dynamics lead to a stationary state, which can be either a steady state maintained by external driving or a true equilibrium state corresponding to thermalization with a bath. 
Parameter estimation is then performed using the properties of this time-independent state. 

In contrast, non-equilibrium sensing exploits the transient dynamics of the probe before it reaches stationarity, allowing information about the parameter of interest to be extracted from the system’s evolving state. 
This dynamical regime can provide enhanced sensitivity or faster information acquisition, particularly when the parameter influences the system’s decay rates or coherent couplings.

In this work, we focus on the non-equilibrium regime and consider an array of quantum emitters (two-level systems) coupled to a one-dimensional semi-infinite waveguide. We begin our analysis with the simplest case of a single emitter acting as a quantum probe for the waveguide properties and then extend the discussion to multiple emitters, where collective interference effects become significant.
\subsection{Single-qubit probe}
\label{Single-qubit probe}
In this section, we consider an emitter, namely a single two-level system (TLS) as a quantum probe to estimate the effective wave number $k_{\mathrm{1D}}$ of the waveguide. We assume that the emitter is initially prepared in its excited state, $\ket{\rm e}$. Under the master equation given in Eq.~\eqref{eq:master}, the time evolution of the emitter’s density matrix can be obtained analytically:
\begin{align}
\rho(t) = e^{-\Gamma t} \, \ket{\rm e}\bra{\rm e} + \big(1 - e^{-\Gamma t}\big) \, \ket{\rm g}\bra{\rm g},\label{single_qubit_dynamics}
\end{align}
where $\Gamma = \Gamma_{0}+\gamma$, in which $\Gamma_{0}=\gamma_{\mathrm{1D}} \left[ 1 - \cos(2 k_{\mathrm{1D}} z) \right]$ and $\gamma$ denote the decay rates into guided and non-guided modes, respectively. Using~\eqref{single_qubit_dynamics} as the state of the quantum probe for the parameter $k_{1\mathrm{D}}$, the corresponding QFI reads
\begin{equation}
F_Q(t) =
\frac{4\,(\gamma_{1\mathrm{D}} z t)^2\, e^{-\Gamma t}\, 
\sin^2(2k_{1\mathrm{D}} z)}{1 - e^{-\Gamma t}}.\label{qfi_single_qubit}
\end{equation}
The $t^2$ term in the numerator reflects the initial quadratic growth of information with interaction time, typical of short-time quantum parameter estimation. 
However, the exponential decay factor $e^{-\Gamma t}$, arising from spontaneous emission into the waveguide and free space, suppresses the QFI at long times as the excited-state population vanishes and hence $\lim_{t \to \infty} F_Q(t) = 0$. 
As a result, $F_Q(t)$ exhibits a non-monotonic time dependence, where it first increases as information about $k_{1\mathrm{D}}$ is imprinted in the emitter’s dynamics, reaches a maximum at an intermediate time, and then decreases as the system approaches the ground state. This transient behavior highlights the inherently non-equilibrium nature of the sensing process, where optimal sensitivity is achieved before relaxation process entirely degrades the encoded information.

Here, we should emphasize that the present results rely on the standard Markov approximation commonly employed in waveguide-QED systems. In this regime, retardation effects can be neglected provided that the photon propagation time remains much shorter than the characteristic timescale of the emitter dynamics~\cite{albrecht2019subradiant,fang2015waveguide}. More specifically, the relevant retardation timescale is given by $\tau_{\rm ret}=L/v_g$, where $L$ denotes the characteristic propagation length of photons in the system and $v_g$ is the group velocity of the guided photons. The Markov approximation is therefore valid when $\tau_{\rm ret} \ll \gamma_{1D}^{-1}$, such that photon-mediated interactions act effectively instantaneously compared to the emitter relaxation dynamics. In the particular case of the single-emitter configuration discussed here, the relevant propagation process corresponds to photons emitted toward the mirror and reflected back to the emitter, leading to a round-trip retardation time $\tau_{\rm fb}\sim 2z/v_g$. Consequently, the quadratic scaling $F_Q\propto z^2$ obtained in Eq.~\eqref{qfi_single_qubit} should be understood within the Markovian regime where $\tau_{\rm fb}\ll \gamma_{1D}^{-1}$. Beyond this regime, delayed coherent feedback and memory effects may modify the transient dynamics and generate additional oscillatory or revival-like features in the QFI evolution.

To establish a reference case, we set the emitter position to $z = d = 1$. 
This choice is not necessarily optimal for maximizing sensitivity, but it provides a convenient baseline for comparison. 
Later, we will explore the effect of spatial randomness, where the emitters are distributed non-uniformly along the waveguide, compared to the case of evenly spaced emitters separated by a fixed distance $d$. 
Here, we first compare the QFI for the cases $z = d$ and $z \neq d$. Fig.~\ref{single_qubit}a illustrates the strong sensitivity of QFI to the emitter position, where even a slight displacement from $z=1$ to $z=1.1$ leads to a substantial enhancement in both sensitivity of the probe and the duration over which the QFI remains non-vanishing. The latter effect arises from the fact that the effective waveguide-induced decay rate $\Gamma_0$ depends on the emitter position and can be suppressed through a reasonable choice of $z$, thereby prolonging the time over which information about $k_{1\mathrm{D}}$ is retained in the probe. To gain further insight into the role of the emitter position, we analyze the maximum of QFI with respect to time. From Eq.~\eqref{qfi_single_qubit}, the QFI attains its maximum value as
\begin{equation}
F_{Q,\max} = \alpha\, \left(\frac{2\ \gamma_{1D}\ z}{\Gamma} \right)^{2} \sin^2(2\ k_{1D}\ z),\label{single_qubit_qfi_max}
\end{equation}
at the optimal time instant
\begin{equation}
t_{\max} = \, \frac{\beta}{\Gamma},\label{time_single_qubit_qfi_max}
\end{equation}
where $\alpha\approx 0.65$ and $\beta\approx 1.6$. 
These expressions show that both the achievable sensitivity and the optimal probing time can be modulated through the emitter position $z$.
Specifically, when $k_{1D}\ z\to \ m\ \pi$, the waveguide-mediated decay $\Gamma_{0}$ is strongly suppressed, and the total dissipation is then dominated by the local, independent decay channel with rate $\gamma$, leading to significantly reduced dissipation. Consequently, QFI remains high and decays much more slowly. This effect is also evident from Eqs.~\eqref{single_qubit_qfi_max} and~\eqref{time_single_qubit_qfi_max}, where both $F_{Q,\max}$ and its characteristic time $t_{\max}$ increase markedly near these positions, demonstrating that careful qubit positioning can significantly enhance the longevity of information in the system. Fig.~\ref{single_qubit}b shows the dependence of $F_{Q,\max}$ on the position of the emitter. The resulting curve displays a pronounced periodicity together with a characteristic double-peak structure within each period. This feature originates from the spatial mismatch between the maxima of the coherent contribution $z^2 \sin^{2} (2 k_{1D} z)$ and the minima of the decay rate $\Gamma_0$. While the former is maximized near the zeros of $\cos^{2}(2k_{1D}z)$, the latter attains its minimum at positions where $\cos^{2}(2k_{1D}z)=1$. As a result, the QFI becomes enhanced either through an increased coherent growth or through a strong suppression of decay, thereby producing two optimal positions per spatial period. Additionally, the height of the peaks also increases with $z$, consistent with the explicit $z^2$ scaling in Eq.~\eqref{single_qubit_qfi_max}. We note that, although the maximal QFI is achieved at specific optimal emitter positions, the enhancement is not restricted to isolated points in space. Owing to the periodic structure displayed in Fig.~\ref{single_qubit}b, neighboring emitter positions can still support comparatively large QFI values, even in the presence of moderate spatial deviations from the optimal configurations. Moreover, both the height and the width of the enhanced QFI regions increase with the emitter--mirror distance, indicating an increased tolerance against positioning imperfections at larger separations. These observations suggest that the sensing enhancement mechanism remains qualitatively robust against moderate spatial uncertainties in experimentally realistic implementations. Later, we further investigate the robustness of the results against imperfect spatial control by considering disordered emitter configurations.

On the other hand, to quantify the temporal window during which the probe retains information about $k_{1D}$, we define a threshold value $\varepsilon$ below which the QFI is considered negligible. Introducing the dimensionless variable $y \equiv \Gamma t$ and defining
\begin{equation}
A = \frac{4(\gamma_{1D}z)^2 \sin^2(2k_{1D}z)}{\Gamma^2},\label{A}
\end{equation}
\begin{figure*}[t!]
\begin{center}
\includegraphics[width=\textwidth]{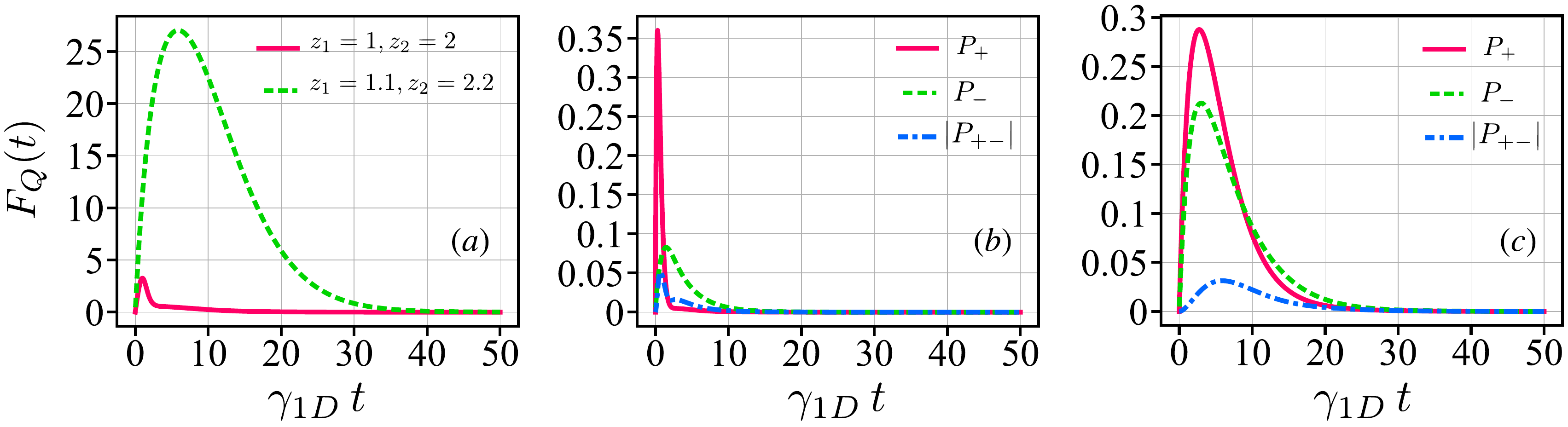}
\end{center}
\caption{Time evolution of QFI with two emitters as the quantum probe, and population and coherence in single excitation subspace for (b) $z_1=1, z_2=2$ and (c) $z_1=1.1, z_2=2.2$. The other parameters are the same as Fig.~\ref{single_qubit}.}
\label{two_qubit_dynamics}
\end{figure*}
the QFI can be expressed as
\begin{equation}
F_Q(t_\varepsilon) = \frac{A\, y_\varepsilon^2 e^{-y_\varepsilon}}{1 - e^{-y_\varepsilon}} = \varepsilon.\label{qfi_single_epsilon}
\end{equation}
Solving this equation for $y$ yields $y_\varepsilon$, from which the characteristic time at which the QFI effectively vanishes can be obtained as
\begin{equation}
t_\varepsilon = \frac{y_\varepsilon}{\Gamma}.\label{time_epsilon}
\end{equation}
We plot this quantity in Fig.~\ref{single_qubit}c, where analogous to $F_{Q,\max}$, a periodic double-peak pattern emerges, whose height grows with $z$, highlighting the crucial role of emitter position in extending the temporal robustness of the probe’s sensitivity.
\subsection{Two-Qubit Probe}
\label{Two-Qubit Probe}
We now consider a two-qubit probe to estimate $k_{1D}$. The dynamics of the two qubits is governed by the master equation \eqref{eq:master}
which embeds both local and collective decays into the free space and the waveguide. To simplify the analysis, we express the dissipative dynamics in terms of the collective jump operators corresponding to the eigenvectors of the decay matrix $\Gamma_{ij}=\Gamma^{\rm s}_{ij}+\gamma \delta_{ij}$. In this representation, each collective mode decays independently at the rate given by the corresponding eigenvalue of $\Gamma_{ij}$. For two qubits, the decay matrix has two eigenvalues, $\Gamma_+$ and $\Gamma_-$, corresponding to the \emph{superradiant} and \emph{subradiant} decay channels, respectively. Diagonalizing $\Gamma_{ij}$ allows us to rewrite the dissipation part as
\begin{equation}
\sum_{\mu}  L_\mu \rho L_\mu^\dagger - \frac{1}{2} \{ L_\mu^\dagger L_\mu, \rho \},\label{dissipation_part_with_collective}
\end{equation}
in which the collective jump operators are defined as
\begin{equation}
L_\mu = \sqrt{\Gamma_\mu}\ \sum_{j=1}^{2} u^j_{\mu} \sigma_j^-, \quad \mu = +,- \label{L_mu}
\end{equation}
where $u^j_{\mu}$ are the components of the eigenvectors of $\Gamma_{ij}$. Using this form of the master equation, the population dynamics becomes separated from the single-excitation subspace. Initializing the qubits in $\ket{\rm e\, e}$, the probability of this double-excitation state decays as
\begin{equation}
P_{ee}(t) = e^{-(\Gamma_+ + \Gamma_-) t}, \label{P_ee(t)}
\end{equation}
while for the single-excitation subspace, one obtains
\begin{equation}
\mathbf{P}^{(1)}(t) = e^{M t} \mathbf{P}^{(1)}(0) + \int_{0}^{t} e^{M (t-s)} \mathbf{f}(s)\, ds,\label{differential_equations_for_P1}
\end{equation}
where $\mathbf{P}^{(1)}(t)=\big(P_{eg}(t)\ P_{c}(t)\ P^{*}_{c}(t)\ P_{ge}(t) \big)^{T}$, where $P_{eg}$, $P_{ge}$ and $P_{c}$ being the population of $\ket{\rm e\, g}$, $\ket{\rm g\, e}$, and their coherence, respectively, while $M$ is the dynamical coefficient matrix and $\mathbf{f}(t)$ shows the feeding vector from the double-excitation state (see Appendix for details).

By defining the single-excitation basis $\ket{S}=\ket{\rm e\, g}$ and $\ket{R}=\ket{\rm g\, e}$, the collective modes can be expressed as
\begin{equation}
\ket{\mu} = u_1^{\mu} \ket{S} + u_2^{\mu} \ket{R}.\label{collective_modes}
\end{equation}
It follows that the corresponding collective decay rates are $\Gamma_{+}=\gamma + \gamma_{1D}\, \big(2 - \cos(2 k_{1D} z_{1}) -\cos(2 k_{1D} z_{2})  \big)$ and $\Gamma_{-}=\gamma$. This indicates that in the absence of local individual decay ($\gamma=0$), the subradiant mode $\ket{-}$ is indeed a \textit{dark state}, immune to direct radiative decay into the waveguide. Nevertheless, its population may still evolve through coherent Hamiltonian dynamics, which can induce transitions via the coherent coupling between the $\ket{S}$ and $\ket{R}$.  
In particular, the off-diagonal elements of the single-excitation block of the density matrix in the collective basis represent the coherence between $\ket{S}$ and $\ket{R}$, which mediates population transfer and in turn, effective decay via $\Gamma_{+}$. Accordingly, we consider the populations within the single-excitation subspace expressed in the collective basis $\{\ket{+},\ket{-} \}$ defined as
\begin{equation}
P_{\mu\nu} = \bra{\mu}  P^{(1)} \ket{\nu},
\label{collective_population}
\end{equation}
where $P^{(1)}$ is two-dimensional state of the single excitation block.

To elucidate the mechanism behind the position-dependent enhancement of metrological performance in the two-qubit probe, we consider the populations of the collective modes and their coherence, as defined in Eq.~\eqref{collective_population}, alongside the QFI. As illustrated in Fig.~\ref{two_qubit_dynamics}a, tuning the emitter positions from $(z_{1},z_{2})=(1,2)$ to $(1.1,2.2)$ significantly enhances the QFI, both in peak value and in temporal persistence. As already observed in the single-qubit probe, the metrological response is generically periodic in position, because the collective decay rate also inherits periodicity with respect to the positions of the emitters. Therefore even small positional shifts can significantly reshape the radiative landscape experienced by the probe, and configurations that suppress rapid decay lead to larger growth and longer persistence of QFI, facilitating non-equilibrium sensing before it eventually vanishes. For the first configuration, where the QFI decays rapidly, the populations and coherence of the collective eigenmodes all relax quickly to zero, as shown in Fig.~\ref{two_qubit_dynamics}b. This behavior reflects dynamics dominated by the fast superradiant channel, with a large $\Gamma_{+}$ that rapidly depletes both populations and coherence in the single-excitation subspace. By contrast, for $(z_1,z_2)=(1.1,2.2)$, the collective populations and their coherence remain substantial over a much longer time (Fig.~\ref{two_qubit_dynamics}c), consistent with a suppressed superradiant rate, and hence enhanced sensitivity over longer duration.

\begin{figure*}[t!]
\begin{center}
\includegraphics[width=\textwidth]{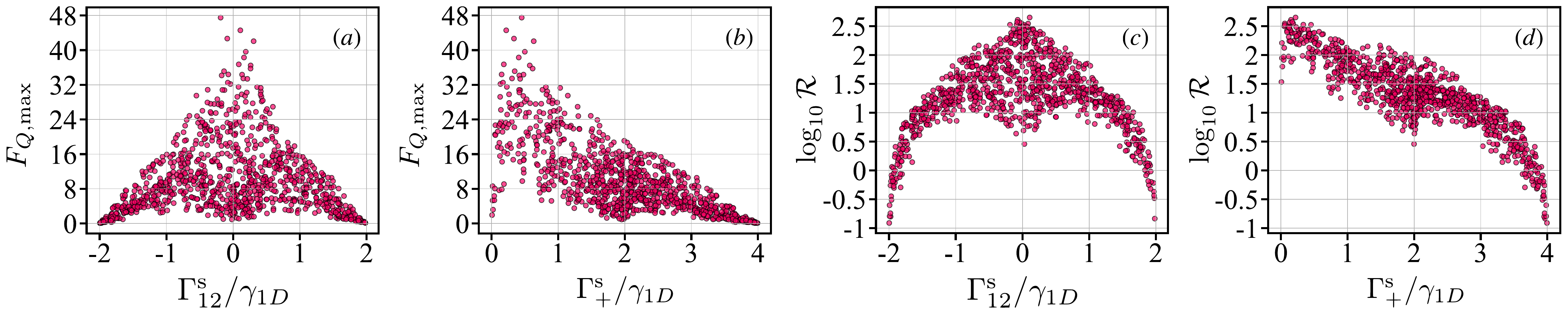}
\end{center}
\caption{Statistical behavior of (a), (b) maximum QFI and (c), (d) durability of information against cross decay term $\Gamma^{\rm s}_{12}$ and superradiant decay rate $\Gamma^{\rm s}_{+}$, for $10^3$ samples of random positions for the two emitters. The other parameters are the same as Fig.~\ref{single_qubit}.}
\label{random_scatter}
\end{figure*}

To shed more light into this observation, we now examine how the maximal QFI behaves as a function of the waveguide-induced cross decay rate $\Gamma^{\rm s}_{12}/\gamma_{1D}=\cos(k_{1D} \lvert z_{1}-z_{2} \rvert)-\cos(k_{1D} \lvert z_{1}+z_{2} \rvert)$. To this end, we include disorder and randomize the emitter positions according to $z_{i}=(i+\delta_{i})d$, where each $\delta_{i}$ is independently sampled from a uniform distribution $\left[-0.5,0.5\right]$.  In Fig.~\ref{random_scatter}a we display $F_{Q,\max}$ against $\Gamma^{\rm s}_{12}/\gamma_{1D}$ for $10^{3}$ disorder realizations. The resulting scatter plot shows a characteristic triangular structure that is symmetrically distributed around $\Gamma^{\rm s}_{12}/\gamma_{1D}=0$, consistent with the fact that $\Gamma^{\rm s}_{12}/\gamma_{1D}$ can take both positive and negative values depending on the relative phase accumulated between emitters. Remarkably, the largest QFI values are obtained in the vicinity of $\Gamma^{\rm s}_{12}/\gamma_{1D}=0$. This regime corresponds to emitter separations satisfying $\lvert z_{1}-z_{2} \rvert=\lvert z_{1}+z_{2} \rvert+ 2 n \pi/k_{1D}$ which simultaneously forces the waveguide-induced coherent coupling $J_{12} \propto \sin(k_{1D} \lvert z_{1}-z_{2} \rvert)-\sin(k_{1D} \lvert z_{1}+z_{2} \rvert)$ to vanish. This leaves only local dissipation channels—the guided-mode decay rates $\Gamma^{\rm s}_{ii}/\gamma_{1D}=1-\cos(2 k_{1D} z_{i})$ and the nonguided decay $\gamma$. In this situation, the collective decay channel is absent, and superradiant emission is suppressed, which manifests as as pronounced peaks in the maximal QFI around $\Gamma^{\rm s}_{12}/\gamma_{1D}=0$. It is worth noting that the above interference mechanism originates from the interplay between photons emitted directly into the waveguide and those reflected back by the mirror at the waveguide boundary. In the present study, we assume an ideal perfectly reflecting mirror in order to clearly isolate the role of these geometry-controlled interference effects. In a more realistic setting, partial reflectivity or additional mirror losses would modify the reflected guided field and consequently alter the interference processes underlying the collective coherent and dissipative couplings, given by Eq.~\eqref{coupling} and~\eqref{decay}. Naturally, since imperfect reflection weakens the reflected component of the guided photons, the mirror-induced interference effects are generally expected to become less pronounced, which may quantitatively affect the suppression of superradiant decay and the associated QFI enhancement. Nevertheless, the underlying interference-based mechanism should remain qualitatively robust provided the reflectivity remains sufficiently high.

In Fig.~\ref{random_scatter}b, we plot $F_{Q,\max}$ against $\Gamma^{\rm s}_{+}/\gamma_{1D}=2-\cos(2 k_{1D} z_{1}) - \cos(2 k_{1D} z_{2} )$ (we exclude $\gamma$ as it is position independent). As shown, large values of $\Gamma^{\rm s}_{+}/\gamma_{1D}$ strongly suppress the maximal QFI, reflecting the rapid depletion of single-excitation population through the fast superradiant channel. In contrast, small $\Gamma^{\rm s}_{+}/\gamma_{1D}$ can allow substantial enhancement of $F_{Q,\max}$, though it does not guarantee a large QFI since the local waveguide-mediated decays $\Gamma^{\rm s}_{ii}$ which depend on the specific positional configuration of the emitters, can still be significant and reduce the QFI.
To complement the analysis based on the peak values of the QFI, we also examine its durability by computing the time-integrated QFI
\begin{equation}
\mathcal{R}=\int F_Q(t)\ dt,\label{int_qfi}
\end{equation}
which measures both magnitude of QFI and the total temporal window during which the probe retains information about $k_{1D}$. As shown in Fig.~\ref{random_scatter}c, $\mathcal{R}$ exhibits an almost “moon-like” distribution, where the largest integrated QFI values are again concentrated around $\Gamma^{\rm s}_{12}/\gamma_{1D}=0$, while thin tails emerge at large large $\lvert \Gamma^{\rm s}_{12} \rvert$ in which $\mathcal{R}$ rapidly diminishes. This is expected since strong $\lvert \Gamma^{\rm s}_{12} \rvert$ increases the superradiant decay rate, as discussed earlier, which limits the longevity of QFI as also shown in Fig.~\ref{random_scatter}d, in which $\mathcal{R}$ is monotonically reduced as $\Gamma^{\rm s}_{+}$ grows.
\subsection{Scaling with the number of emitters}
\label{Scaling with the number of emitters}
In this section, we extend our analysis to the case of $N$-body probes and investigate how metrological sensitivity emerges dynamically and scales with the number of emitters. A central element of this extension is the spatial arrangement of emitters. The two-body coherent coupling and decay between each two emitters, given in Eq.~\eqref{coupling} and~\eqref{decay}, can be rewritten in the compact forms $J_{ij}/\gamma_{1D}=-\cos(k z_>) \sin(k z_<)$ and $\Gamma^{\rm s}_{ij}/\gamma_{1D}=2\sin(k z_>) \sin(k z_<)$, respectively, where $z_>=\max\{z_i,z_j\}$ and $z_<=\min\{z_i,z_j\}$. Importantly, while it is possible to minimize the collective decay $\Gamma^{\rm s}_{ij}$ for a specific emitter pair, the relative ordering of emitters means that $z_i$ and $z_j$ switch roles for different pairs. As a result, it is impossible to simultaneously minimize all pair of $\Gamma^{\rm s}_{ij}$, while maintaining large $J_{ij}$ across the entire array, and therefore the exact optimal configuration cannot be achieved. A full numerical optimization of all emitter positions is also intractable for larger $N$, since optimizing over $N$ independent position coordinates quickly becomes numerically demanding. For this reason, we pursue specific representative and experimentally motivated positioning strategies. We first consider the simple choice where $z_{i}=i\, d$ corresponding to uniform integer spacing where $i=1,2,\ldots,N$ and $d$ is a constant. As shown previously in the single- and two-emitter cases, such regular spacing generally leads to rapid collective decay due to constructive interference, and therefore does not yield particularly large sensitivity. We then consider a slightly detuned linear spacing of the form $z_i = (i + i\,\Delta)\, d$ in which $\Delta=0.1$. As previously observed, even modest relative shifts of the emitters can substantially suppress the dominant superradiant decay channel, resulting in enhancement of QFI. Although this configuration is not guaranteed to be fully optimal, it consistently outperforms the simple uniform integer spacing. In what follows, we refer to these positioning strategies as the uniform, shifted configurations, respectively. Later, we also return to the case of disordered positioning, where the emitters are randomly spaced. Besides providing a broader class of spatial configurations, this scenario also allows us to examine the robustness of the metrological enhancement against imperfections in emitter positioning and deviations from ideally engineered geometries, which are naturally expected in realistic experimental implementations. At this stage, however, we restrict our analysis to the two mentioned configurations.
\begin{figure*}[t!]
\begin{center}
\includegraphics[width=\textwidth]{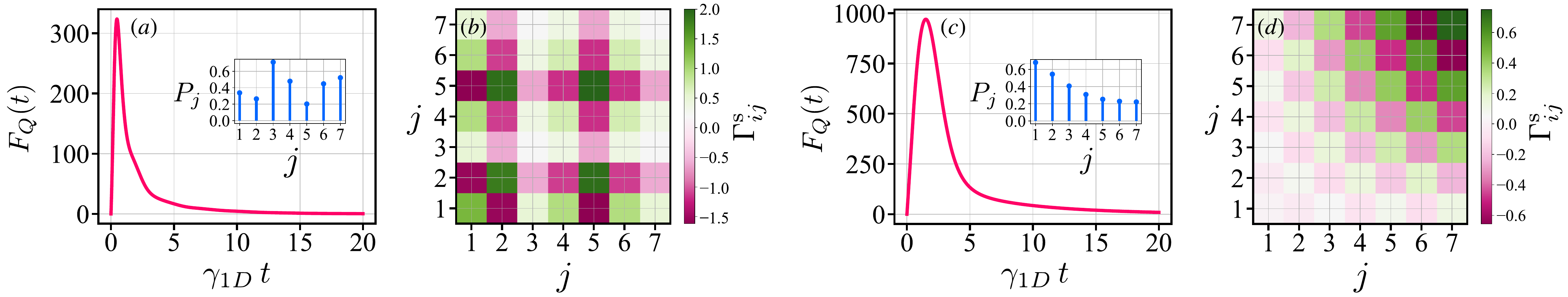}
\end{center}
\caption{(a), (c) Time evolution of QFI with 7 emitters as a quantum probe for (a) uniform and (c) shifted position configurations. The insets show the local excitation probabilities at the time instant for which the QFI reaches its maximum, (b), (c) the waveguide-induced decay rate $\Gamma^{\rm s}_{ij}$ for corresponding positioning strategies. The parameters are the same as Fig.~\ref{single_qubit}.}
\label{seven_qubit}
\end{figure*}
Fig.~\ref{seven_qubit}a and~\ref{seven_qubit}c depict the time evolution of the QFI in a seven-qubit array for the uniform and shifted emitter configurations, respectively, with the latter arrangement producing markedly higher and more persistent sensitivity, consistent with the behavior seen in smaller probes. The insets also display the local excitation probabilities of each emitter at the time when the QFI reaches its maximum, given by $P_{j}(t_{\rm max})=\bra{{\rm e}_{j}}\rho(t_{\rm max}) \ket{{\rm e}_{j}}$, where $\ket{{\rm e}_{j}}$ is the excited state of the j-th emitter, plotted as a function of the emitter index $j$. For the uniform case, $P_{j}(t_{\rm max})$ show no clear spatial pattern and different emitters lose excitation at different rates, but in an irregular and non-monotonic fashion. This behavior is fully consistent with the structure of the waveguide-induced decay matrix $\Gamma^{\rm s}_{ij}$, as shown in Fig.~\ref{seven_qubit}b. In this geometry, both the local and the collective decay channels tend to be largest around some interior sites (e.g., $j=2,5$). This decay landscape effectively distributes dissipation throughout the array without regularly privileging any spatial direction. In stark contrast, the shifted configuration produces a highly ordered excitation profile, where the probabilities decay monotonically from the leftmost emitter, near the mirror, to the rightmost emitters, nearest to the escape region of the waveguide, such that $P_{j}(t_{\rm max})>P_{j+1}(t_{\rm max})$. This reflects a strongly directionalized excitation flow enforced by the waveguide. The corresponding decay matrix (Fig.~\ref{seven_qubit} d) clearly reveals this behavior, where both $\Gamma^{\rm s}_{ii}$ and $\Gamma^{\rm s}_{ij}$ ($i \neq j$) increase steadily toward the right end of the array, thus, dissipation is highly biased in one direction, which in turn, can generate phase gradient and coherence between adjacent emitters, leading to a more responsive collective state and consequently higher sensitivity. 

This behavior further highlights that, in the present setting, the spatial arrangement of emitters itself acts as the primary control mechanism of the sensing protocol. Since both the coherent couplings $J_{ij}$ and the collective decay rates $\Gamma^{\rm s}_{ij}$ are directly determined by the emitter positions, modifying the inter-emitter spacing effectively reshapes the interference landscape experienced by the guided photons. Physically, this originates from constructive or destructive interference between photons emitted by different emitters and reflected by the mirror, which determines whether the collective radiative channels are enhanced or suppressed. Consequently, particular geometrical configurations can either suppress or enhance collective decay pathways, including the superradiant channels, thereby stabilizing populations and coherences that carry metrological information. In this sense, the emitter geometry acts as a passive resource for controlling sensitivity, rather than conventional metrological schemes that rely on external driving fields, dynamical modulation, or engineered control pulses to manipulate the probe~\cite{pang2017optimal,liu2020quantum,mishra2021driving,mukherjee2019enhanced}. 

Importantly, such spatial engineering is experimentally feasible in several waveguide-QED platforms, including superconducting-qubit arrays coupled to transmission lines and atoms trapped near photonic crystal waveguides, where emitter separations and geometry-dependent collective radiative effects can be controlled with high precision~\cite{brehm2021waveguide,goban2015superradiance,kannan2020waveguide,lodahl2004controlling}. In these architectures, uniformly spaced emitter arrays naturally arise from lithographically patterned superconducting circuits or periodically trapped atomic arrays. Likewise, the shifted configurations considered here can be implemented through controlled variations of the emitter spacing during fabrication or trapping, while disordered arrangements may be engineered intentionally or can emerge from residual fabrication imperfections and stochastic trapping variations.

We further emphasize that the arrays considered in the present work contain at most seven emitters and therefore remain within a relatively compact regime. For substantially larger arrays with increased effective waveguide length, however, finite photon propagation times may become non-negligible and generate stronger retardation-induced memory effects beyond the Markovian regime~\cite{tufarelli2014non,dinc2019exact} which may modify the transient dynamics through delayed photon-mediated feedback between emitters. Consequently, the QFI profile could develop additional oscillatory or revival-like features, and the temporal interval over which the QFI remains nonzero may become prolonged due to partial backflow of information from the waveguide to the emitters. Nevertheless, the present work focuses on the Markovian regime, which already captures the essential geometry-controlled enhancement mechanism.

Next, we examine how the metrological sensitivity scales with the number of emitters, $N$, coupled to the waveguide. We consider both the maximum of QFI $F_{Q,\max}$ and its durability $R$, previously introduced as a measure of the temporal robustness of the metrological response. In addition to the uniform and shifted configurations, we also consider the disordered configuration with randomized emitter positions of the form $z_i = (i + \delta_i)\, d$ where $\delta_{i} \in \left[-0.5,0.5 \right]$ randomly drawn from a uniform distribution. For this case, the reported quantities are obtained by performing an ensemble average over independent disorder realizations. Specifically, for each realization we compute the full time-dependent quantum Fisher information $F_Q(t)$, and the disorder-averaged result is then obtained as $ \langle F_Q(t) \rangle = \frac{1}{M}\sum_{m=1}^{M} F_Q^{(m)}(t)$, where $M$ denotes the number of disorder realizations. In the present simulations, convergence is ensured by using a sufficiently large ensemble of realizations.
. It is worth of mentioning that the localization effects are not addressed here since prior works in many-body waveguide indicate that localization requires that at most half of the emitters are excited~\cite{fayard2021many}. In our case, however, all emitters start fully excited, and the system size is also limited, precluding a meaningful analysis of many-body localization.

\begin{figure}[h]
\begin{center}
\includegraphics[width=\columnwidth]{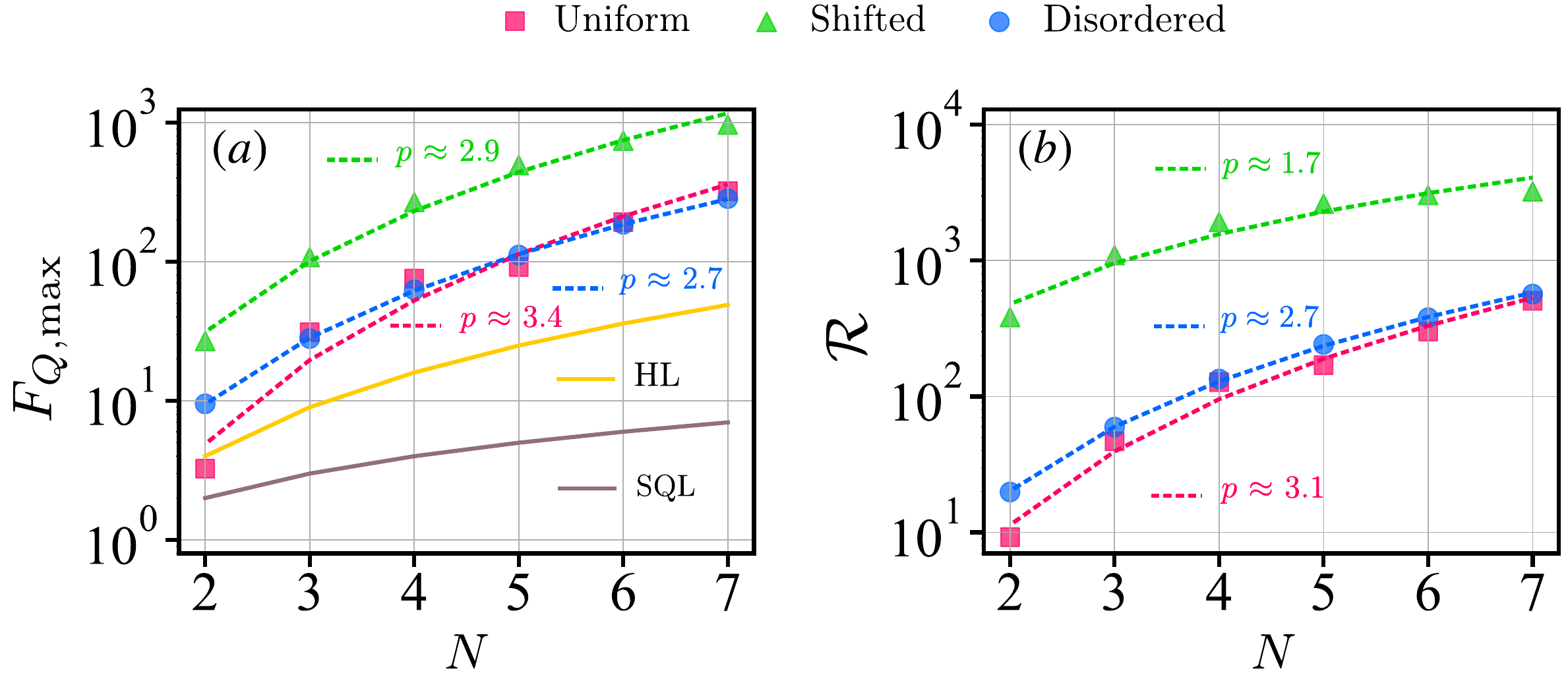}
\end{center}
\caption{Scaling of (a) maximum QFI and (b) its durability with the number of emitters embedded in our quantum probe. The dashed lines show the fitted curves corresponding to each positioning strategy, proportional to $N^{p}$. In panel (a), we have also included the scaling associated to standard quantum limit (SQL) and Heisenberg limit (HL), for reference. For the disordered case, the results are averaged over $10^{3}$ number of disorder realization for all $N$, except for $N=7$ for which the number of sample is $4 \times 10^{2}$ The other parameters are the same as Fig.~\ref{single_qubit}.}
\label{scaling_with_N}
\end{figure}

Fig.~\ref{scaling_with_N}a illustrates how the maximum of QFI scales with the number of emitters for the considered configuration discussed above. For each case, we additionally perform a power-law fitting of the form $F_{\rm fit} = A\, N^{p}$ which allows us to quantitatively compare the observed scaling with the Standard Quantum Limit (SQL) and the Heisenberg Limit (HL), for which the QFI scales as $F_{Q} \propto N$ and $F_{Q} \propto N^{2}$, respectively. As evident from the figure, all positioning strategies exceed both bounds by a wide margin, thereby indicating pronounced super-Heisenberg scaling. Among the three strategies, the shifted configuration yields the largest values of $F_{Q,\max}$ across the accessible range of system sizes. The uniform and disordered configurations remain comparatively close. For small $N$ the disordered arrangement provides a slightly higher $F_{Q,\max}$, whereas for larger number of emitters, the uniform configuration eventually can become dominant. The modest fluctuations observed in the disordered case stem from the inherent sample-to-sample variability of the randomized emitter positions. The fitted exponents and prefactors are $p\approx(3.4,2.9,2.7)$ and $A\approx(0.45,4.2,1.4)$ for the uniform, shifted, and disordered configurations, respectively. These values reveal that, although the shifted configuration gives the largest absolute magnitude of sensitivity over the considered number of emitters, the uniform configuration exhibits the steepest asymptotic power-law growth with $N$. This reflects the fact that the shifted configuration more effectively suppresses collective decay for small and intermediate system sizes, while the uniform spacing leads to a structured interference pattern that produces a stronger scaling exponent, despite its smaller prefactor. Importantly, the persistence of super-Heisenberg scaling even in the disordered configuration indicates that the geometry-induced sensing enhancement is not restricted to perfectly engineered emitter arrays, but remains qualitatively robust against moderate spatial uncertainties and imperfections in emitter positioning.

It is useful to recall that, in non-equilibrium quantum sensing, both the interrogation time and the system size may in general act as relevant metrological resources, with the QFI often exhibiting a scaling behavior of the form $F_Q \sim t^x N^y$. Within this framework, the standard quantum limit corresponds to $x=y=1$, while $x=y=2$ defines the Heisenberg limit~\cite{montenegro2025quantum,ilias2022criticality}. In this work, the scaling of the optimized maximal QFI with the number of emitters already exhibits pronounced super-Heisenberg behavior, as evidenced by the fitted exponents $p=y>2$ for all considered positioning strategies. Regarding the temporal dependence, the short-time evolution of the QFI displays the expected quadratic behavior in time, $F_Q\propto t^2$, as explicitly shown for the single-emitter case in Eq.~\eqref{qfi_single_qubit}.

Importantly, the observed enhancement with increasing system size is not achieved at the expense of longer interrogation times. By comparing the time-dependent dynamics shown in Figs.~\ref{two_qubit_dynamics}a and~\ref{seven_qubit}a,c, one finds that the optimal interrogation time at which the QFI reaches its maximum does not increase with the number of emitters and, in fact, becomes shorter for larger arrays. This behavior originates from stronger collective interference effects, which accelerate the buildup of metrological information while simultaneously enhancing the maximal achievable sensitivity. Consequently, the super-Heisenberg scaling observed here arises genuinely from geometry-controlled many-body effects rather than from consuming additional temporal resources.

We further note that, in some quantum sensing protocols, it is useful to normalize the QFI by the required interrogation time in order to account explicitly for time as a metrological resource~\cite{rossi2020noisy,albarelli2018restoring}. Nevertheless, in the present case the optimal interrogation times remain relatively short compared to the substantial enhancement of the maximal QFI with system size. One therefore expects the observed super-Heisenberg scaling to remain qualitatively robust even under such time-normalized considerations.

One should note that in our study, since all emitters are initialized in the fully excited state, the corresponding Hilbert space has to be fully retained without truncation. Thus, extending the analysis to substantially larger system sizes becomes numerically demanding. This challenge is particularly pronounced, in the disordered configuration, where each data point requires averaging over many independent realizations of the random positions. Consequently, the precise numerical values of the fitted exponents may evolve as $N$ grows beyond the range accessible here. Although the present results clearly demonstrate super-Heisenberg scaling within the numerically accessible regime, the ultimate asymptotic behavior for substantially larger emitter arrays remains an open question. In particular, as the number of emitters grows, the increasingly complex interplay between collective interference effects, coherent exchange interactions, and dissipative decay channels may eventually modify the observed scaling behavior or lead to crossover and partial saturation effects. Indeed, while the fitted curves for the maximal QFI suggest that the sensitivity continues to increase rapidly with the number of emitters, their progressively reduced slope at larger $N$ also indicates a tendency toward gradual saturation. Nevertheless, within the range of system sizes reliably accessible in the present work, all considered positioning strategies exhibit clear scaling behavior beyond the Heisenberg limit.

Similarly, in Fig.~\ref{scaling_with_N}b we examine the scaling of $\mathcal{R}$ . The overall trends and behavior closely resemble those observed for $F_{Q,\max}$, with the exception that the disordered configuration now slightly outperforms the uniform arrangement across the entire range of considered $N$. Nonetheless, the two curves gradually converge for larger system sizes. These results indicate that both the sensitivity of the probe and its temporal persistence can scale with the number of embedded emitters, a feature that is crucial for non-equilibrium sensing.

We should emphasize that the present analysis assumes the standard waveguide-QED regime in which the guided modes exhibit approximately linear dispersion around the emitter resonance frequency~\cite{lalumiere2013input,sheremet2023waveguide}. Under this approximation, the photon-mediated coherent and dissipative couplings acquire the simple analytic forms given in Eqs.~\eqref{coupling} and~\eqref{decay}, while a single guided channel dominantly mediates the emitter interactions. In more realistic situations involving stronger dispersion or multimode propagation, the effective emitter–emitter couplings can become modified through frequency-dependent propagation phases and additional guided decay channels. Physically, such effects are generally expected to weaken the interference processes responsible for suppressing collective radiative decay, thereby leading to quantitative reductions in the achievable QFI enhancement and modifying the optimal emitter configurations. Nevertheless, provided that a dominant guided mode still governs the dissipative dynamics, the underlying geometry-controlled interference mechanism is expected to remain qualitatively robust.
\section{Conclusion}
\label{Conclusion}
In this work, we have demonstrated that arrays of quantum emitters coupled to a one-dimensional waveguide provide a powerful, intrinsically non-equilibrium platform for precision metrology. By exploiting the transient dynamics of spontaneously emitting probes—rather than relying on steady-state properties—we showed that spatial configuration alone can dramatically enhance quantum Fisher information. For a single emitter, optimal placement relative to the guided mode suppresses waveguide-induced decay, thereby prolonging the temporal window in which information about the wave number is retained. Extending the analysis to two emitters revealed that the enhancement mechanism is governed by the structure and stability of the single-excitation subspace, where appropriately chosen separations suppress superradiant channels, stabilize population and coherence dynamics, and yield significantly larger and longer-lasting sensitivity.

Our statistical study over randomized configurations further underscores the central role of collective decay engineering. Both the maximum QFI and its integrated value over temporal window attain their maximum when the waveguide-mediated cross-decay rate vanishes, which eliminates superradiance and allows the probe to retain information for extended periods. Scaling the probe size further confirms the generality of our findings, where across all positioning strategies—including uniform, shifted, and fully randomized spatial configurations—collective-decay suppression enabled by appropriate emitter arrangements leads to super-Heisenberg scaling of the maximal transient QFI.

Recent experimental progress in waveguide QED—such as the thermometry scheme based on bright and dark collective states of superconducting qubits~\cite{sharafiev2025leveraging}—highlights the growing potential of guided-mode architectures for quantum sensing. Our results complement and extend these developments by showing that, even in the simplest dissipative regimes, geometric control of emitter positions alone can markedly improve metrological performance.

Our results establish transient, geometry-assisted metrology in waveguide QED as a promising route toward efficient quantum sensors, where controllable collective effects and scalable enhancements can be harnessed without relying on highly resourcefull initial states.


\section*{Data Availability Statement}
The data that support the findings of this study are available from the corresponding author
upon reasonable request.
\section*{Conflict of interest}
The authors have no conflicts to disclose.

\appendix

\section{Collective jump operators}
\label{Appendix}
In this appendix, we outline the procedure used to express the master equation in terms of the collective decay channels and to analyze the dynamics within the single-excitation subspace.

The collective decay channels are obtained by diagonalizing the decay-rate matrix $\Gamma_{ij}$ as
\begin{equation}
\Gamma = U \Lambda U^{T},
\end{equation}
where $\Lambda = \mathrm{diag}(\lambda_1, \lambda_2, \dots)$ is the diagonal matrix of the eigenvalues of $\Gamma_{ij}=\Gamma^{\rm s}_{ij}+\gamma \delta_{ij}$, and $U=\big(\mathbf{u}_{1} \ \mathbf{u}_{2} \ \dots \big)$
with the normalized eigenvectors denoted as 
\begin{equation}
\mathbf{u}_{\mu} =
\begin{pmatrix}
u_{\mu}^{1} \\[0.5em]
u_{\mu}^{2} \\[0.5em]
\vdots
\end{pmatrix}.
\end{equation}

The master equation can then be expressed as
\begin{equation}
\dot{\rho} = -i [H, \rho] + \sum_{\mu}  \left( L_\mu \rho L_\mu^\dagger - \frac{1}{2} \{ L_\mu^\dagger L_\mu, \rho \} \right),
\end{equation}
where the collective jump operators take the form
\begin{equation}
L_{\mu} = \sqrt{\Gamma_{\mu}}\, \mathbf{u}_{\mu}^{T} \boldsymbol{\Sigma},
\end{equation}
with
\begin{equation}
\boldsymbol{\Sigma} =
\begin{pmatrix}
\sigma_{1} \\
\sigma_{2} \\
\vdots
\end{pmatrix}.
\end{equation}
In case of a two-qubit system, the eigenvalues are given by
\begin{align}
\Gamma_{+} &= \frac{1}{2} \left(\Gamma_{11} + \Gamma_{22} + \sqrt{4 \Gamma_{12} \Gamma_{21} + (\Gamma_{11} - \Gamma_{22})^2} \right), \\
\Gamma_{-} &= \frac{1}{2} \left(\Gamma_{11} + \Gamma_{22} - \sqrt{4 \Gamma_{12} \Gamma_{21} + (\Gamma_{11} - \Gamma_{22})^2} \right),
\end{align}
and the corresponding normalized eigenvectors are
\begin{align}
\mathbf{u}_{+} &=
\begin{pmatrix}
\dfrac{1}{\sqrt{1 + \left( \dfrac{\Delta_{+}}{\Gamma_{12}} \right)^2}} \\[1.5ex]
\dfrac{\Delta_{+}}{N_{+}}
\end{pmatrix},
\label{u_plus}
\end{align}
\begin{align}
\mathbf{u}_{-} &=
\begin{pmatrix}
\dfrac{1}{\sqrt{1 + \left( \dfrac{\Delta_{-}}{\Gamma_{12}} \right)^2}} \\[1.5ex]
\dfrac{\Delta_{-}}{N_{-}}
\end{pmatrix},
\label{u_minus}
\end{align}
where $\Delta_{\pm}=\Gamma_{\pm}-\Gamma_{22}$ and $N_{\pm}=\sqrt{\Gamma^{2}_{12}+\Delta^{2}_{\pm}}$. The collective jump operators are then defined as
\begin{equation}
L_+ = \sqrt{\Gamma_{+}}\ (u^+_1 \sigma_1 + u^+_2 \sigma_2), \qquad
L_- = \sqrt{\Gamma_{-}} (u^-_1 \sigma_1 + u^-_2 \sigma_2).\label{L_{+,-}}
\end{equation}

Plugging these into the master equation, Eq.~\eqref{eq:master}, and working in the basis $\{\ket{\rm ee}, \ket{\rm e\, g}, \ket{\rm g\, e}, \ket{\rm gg}\}$, one can obtain the time evolution of the density matrix elements. For the double-excitation state, the population $P_{ee}(t)$ satisfies $\dot{P}_{ee}(t) = -(\Gamma_+ + \Gamma_-) P_{ee}(t)$ whose solution is
\begin{equation}
P_{ee}(t) = e^{-(\Gamma_+ + \Gamma_-) t},\label{P_{ee}}
\end{equation}
since $P_{ee}(0)=1$
For the single-excitation subspace, the populations and coherences
\begin{equation}
\mathbf{P}^{(1)}(t) =
\begin{pmatrix}
P_{eg}(t) \\
P_{c}(t) \\
P^{*}_{c}(t) \\
P_{ge}(t)
\end{pmatrix},\label{P}
\end{equation}
satisfy a linear differential equation
\begin{equation}
\dot{\mathbf{P}}^{(1)}(t) = M \mathbf{P}^{(1)}(t) + \mathbf{f}(t),\label{single_excitation_equations}
\end{equation}
where the coefficient matrix $M$ reads
\begin{widetext}
\begin{equation}
M =
\begin{pmatrix}
-\Gamma_{11} & i J_{21} - \Gamma_{21}/2 & -i J_{12} - \Gamma_{12}/2 & 0 \\
i J_{12} - \Gamma_{12}/2 & -i(J_{11}-J_{22}) - (\Gamma_{11}+\Gamma_{22})/2 & 0 & -i J_{12} - \Gamma_{12}/2 \\
-i J_{21} - \Gamma_{21}/2 & 0 & i(J_{11}-J_{22}) - (\Gamma_{11}+\Gamma_{22})/2 & i J_{21} - \Gamma_{21}/2 \\
0 & -i J_{21} - \Gamma_{21}/2 & i J_{12} - \Gamma_{12}/2 & -\Gamma_{22}
\end{pmatrix}
\label{Mmatrix}
\end{equation}
\end{widetext}

and the inhomogeneous feeding term from the double-excitation state is
\begin{equation}
\mathbf{f}(t) =
\begin{pmatrix}
\Gamma_{11} \\
\Gamma_{12} \\
\Gamma_{21} \\
\Gamma_{22}
\end{pmatrix} P_{ee}(t).\label{feeding}
\end{equation}
The set of equations~\eqref{single_excitation_equations} can be solved numerically. Finally, the ground state population is obtained from $P_{gg}(t) = 1 - P_{ee}(t) - P_{eg}(t) - P_{ge}(t)$.

\bibliography{refs}

\end{document}